\documentclass[%
 aip,
 amsmath,amssymb,
 reprint,%
]{revtex4-1}

\usepackage{graphicx}
\usepackage{dcolumn}
\usepackage{bm}

\usepackage[utf8]{inputenc}
\usepackage[T1]{fontenc}
\usepackage{mathptmx}
\usepackage{etoolbox}
\usepackage{xcolor}
\usepackage{subfigure}
\usepackage{caption}
\usepackage{amsmath}
\usepackage{bm}

\makeatletter
\def\@email#1#2{%
 \endgroup
 \patchcmd{\titleblock@produce}
  {\frontmatter@RRAPformat}
  {\frontmatter@RRAPformat{\produce@RRAP{*#1\href{mailto:#2}{#2}}}\frontmatter@RRAPformat}
  {}{}
}%
\makeatother
\begin{document}

\preprint{AIP/123-QED}

\title[LES of rotating stratified flows]{Evaluation of eddy viscosity-based models in decaying rotating stratified turbulence}

\author{Kiran Jadhav}
\affiliation{ Department of Mechanical Engineering, Indian Institute of Technology Bombay, Mumbai, Maharashtra, India, Pincode: 400076}%
\author{Rahul Agrawal}
\affiliation{Department of Mechanical Engineering, Stanford University, Palo Alto, California, United States of America, Zipcode: 94305 }
\author{Abhilash J.\ Chandy}%
\email{achandy@iitb.ac.in}
\affiliation{ Department of Mechanical Engineering, Indian Institute of Technology Bombay, Mumbai, Maharashtra, India, Pincode: 400076}
%


\date{\today}

\begin{abstract}
The results of large eddy simulation (LES) using three sub-grid scale models, namely: constant coefficient Smagorinsky, dynamic Smagorinsky, and a dynamic Clark model, for rotating stratified turbulence in the absence of forcing using large-scale isotropic initial condition, are reported here. The LES results are compared to in-house direct numerical simulation (DNS) 
for establishing grid-independence requirements. Three cases with varying ratios of Brunt-Vaisala frequency to the inertial wave frequency, $\mathcal{N}/f$, have been chosen to evaluate the performance of LES models. The Reynolds number and $\mathcal{N}/f$ are chosen as (a) Case 1: $Re=3704$, $\mathcal{N}/f=5$, (b) Case 2: $Re=6667$, $\mathcal{N}/f=40$ and, (c) Case 3: $Re=6667$, $\mathcal{N}/f=138$. This framework is used to illustrate the relative magnitudes of the stratification and rotation which is observed in geophysical flows. Various quantities including turbulent kinetic energy ($tke$), turbulent potential energy ($tpe$), total dissipation, potential and total energy spectra, and their fluxes, are analyzed to understand the predictive capability of the various LES models. Results show that all the SGS model predictions are very similar, with the classical Smagorinsky model displaying the highest deviation compared to DNS. The effect of an increase in value of $\mathcal{N}/f$ is also seen in the results of LES with an increase in the oscillations observed in the evolution of $tke$ and $tpe$ and a reduction in dissipation.  The spectral analysis shows that the dynamic Clark and Smagorinsky models predict the large-scale physics ($\kappa < 10$), while the small scales ($10< \kappa < 64$) energy is under-predicted.
\end{abstract}

\maketitle

\textbf{Keywords}: Turbulence; LES; rotating and stratified flows; SGS models

\section{Introduction}

Rotating stratified turbulence (RST) is fundamentally observed in geophysical and astrophysical flows \cite{van2008refined,marino2013emergence}, with the vertical scales being smaller in length than the horizontal scales of motion due to the action of the buoyancy and rotation forces. Nondimensionally, the flow regime is characterized by a combination of Reynolds, Froude, and Rossby numbers. Owing to its physical significance, such flows have been studied extensively in the past \cite{HAN02,LIN05,LIE05}. Lindborg \cite{LIN05} considered elongated boxes to study the emergence of a direct energy cascade in RST with a Kolmogorov spectrum in the horizontal direction for a Rossby number greater than $\sim 0.1$. It was suggested that for large ratios of the stratification to rotation, or $\frac{N}{f} > 45 $, the aspect ratio of the computational domain plays an important role in influencing the dynamics of the flow. Marino et al. \cite{MAR15} studied stably stratified flows, and found that crossing between the wave and vortex modes occurs at a smaller scale for the tropospheric (less stable) flows than for the stratospheric (more stable) flows. In another work, Marino et al. \cite{MAR13} showed that within the Boussinesq-assumption framework, helicity (velocity-vorticity correlation), as observed in supercell storms and hurricanes and other geophysical flows is spontaneously created due to an interplay between buoyancy and rotation in large-scale atmospheric and oceanic flows.

Rosenberg et al. \cite{ROS15} analyzed RST for the abyssal southern ocean at mid-latitudes and found that the largest scales were dominated by rotation, with a negative energy flux, and that for scales larger than a critical scale, the constant-flux range is one where the source of the energy is the potential energy stored in the large-scale gravity waves. Rosenberg et al. \cite{ROS16} parameterized the buoyancy Reynolds number from very small values to up to $10^{5}$ and found that neither rotation nor the ratio of the Brunt-Vaisala frequency to the inertial frequency seemed to play a major role in the absence of forcing in the global dynamics of the small-scale kinetic and potential modes. Cambon et al. \cite{CAM04} studied decaying rotating stratified flow for $\frac{N}{f}=0.1$, 1, 10 and found a decrease in diffusivity in the vertical direction as compared to horizontal. Recently, Pouquet et al. \cite{PAM19} performed high-resolution, 3D, direct numerical simulations (DNS) of decaying rotating stratified flows. They confirmed the role played by large-scale intermittency, occurring in vertical velocity and temperature fluctuations in RST, as an adjustment mechanism of the energy transfer in the presence of strong waves.   

For motions slower than the rotation rate, a quasi-2D state has been shown in rotating flows \cite{smith1996crossover,greenspan1968theory}, and it is expected that this effect gets amplified in the presence of buoyancy. 
Rotation re-distributes energy amongst different modes by nonlinear mechanisms \cite{van2008refined,greenspan1968theory}, which breaks the isotropy of vorticity along the rotation axis. Previous studies \cite{10.1175/2009JAS3090.1,10.1175/1520-0469(1986)043<0126:TSEAPO>2.0.CO;2} have shown that helicity is an important quantity in long-living storms through suppression of energy dissipation. Thus analysis of homogeneous-rotating-stratified turbulence with periodic boundary conditions is a good first step in canonical problems to understand the underlying physics. Marino et al. \cite{marino2013emergence} performed a parametric DNS study of helical-rotating-stratified flows, and showed a spontaneous production of helicity at large scales for $\frac{N}{f} > 1$.

The computational costs involved in performing DNS, an approach frequently used while resolving the entire range of scales in practical flows are very high \cite{choi2012grid}. In atmospheric flows, the largest eddies that scale with the boundary layer height are six orders of magnitude larger than the smallest Kolmogorov scales. Since a full resolution of all turbulence scales is only possible for very low Reynolds numbers, more focus has been on a large eddy simulations (LES) methodology, which relies on a subgrid-scale (SGS) model to represent the effects of the unresolved small-scale turbulence. A widely-used SGS model is the classical, constant coefficient Smagorinsky \cite{SAM63} model. Germano et al. \cite{GER91} proposed a dynamic variant of the Smagorinsky model in which the model coefficient is computed within the simulation based on the energy content information between test and grid scales. Several other models have been developed since, and details of those studies can be found in existing literature \cite{ghosal1995dynamic,rozema2015minimum,agrawal2022non}. 

In the past, LES studies have been performed for rotating turbulence. Kobayashi and Shimomura \cite{KOB01} examined various dynamic SGS models such as the dynamic Smagorinsky model (DSM), the dynamic mixed model (DMM), the dynamic Clark model (DCM), and the dynamic two-parameter Clark model (DTCM) for rotating homogeneous decaying turbulence. All models show similar performance and are in good agreement with filtered DNS with the DSM predicting a slightly slower decay of the turbulent kinetic energy than the DNS. Shao et al. \cite{SHAO05} suggested a dynamic subgrid eddy viscosity model based on local volume-integrated filtered velocity structure function equations. This model was able to simulate rotating homogeneous turbulence at a very high rotation rate (with a micro-Rossby number smaller than 0.1). Yang and Domaradzki \cite{YANG04} performed LES of rotating turbulence using the truncated Navier-Stokes (TNS), and observed better agreement compared to excessively dissipative models. Lu et al. \cite{LU07} tested various SGS models, algebraic, gradient, and scale similarity to one-equation viscosity and non-viscosity dynamic structure models, and compared results to $128^{3}$ DNS results for rotating turbulence. They found that the similarity-type consistent dynamic structure model (SCDSM) and gradient-type consistent dynamic structure model (GCDSM) perform well in comparison to scale-similarity, gradient, Smagorinsky, and dynamic Smagorinsky models.

In this work, the problem of homogeneous-decaying-rotating-stratified turbulence is considered and a parametric analysis of the $\frac{N}{f}$ ratio using commonly used SGS models such as the constant coefficient, dynamic Smagorinsky, and the mixed Clark non-linear subgrid-scale models \cite{CRF79} is performed. The performance of these three subgrid-scale models is evaluated against DNS at high Reynolds numbers to establish their predictive capabilities for such flows.  

\section{Formulation}

In LES, the large scales are separated from the small scales based on the filtration operation. The filtered variable (or resolved scales) is denoted $\overline{(...)}$ and is defined as follows \cite{LEONARD1975237}:
\begin{equation}
    \overline{f}(x)=\int_{D} f(x') G(x,x';\Delta) dx'
\end{equation}
where $D$ is the computational domain, $G$ is the filter function, and $\Delta$ is the filter width. The unfiltered parameters (DNS values) like velocity and density are denoted by $u$ and $\rho$. If $\overline{u}$ and $\overline{\rho}$ represent the filtered velocity and density fields respectively, then the non-dimensional governing equations are given by: 
\begin{equation}
{\partial}_{i}\overline{u}_{i}=0,
\end{equation}
\begin{equation}
\partial_{t}\overline{u}_{i}+\overline{u}_j\partial_{j}\overline{u}_{i}=-{\partial}_{i}\overline{p}+\frac{1}{Re_{0}}{\partial}_{jj}\overline{u}_{i}-\frac{1}{Fr_{0}^2}\overline{\rho}\widehat{z}-{\partial}_{j}\tau_{ij} - \frac{1}{Ro_0} \hat{z} \times \overline{u}_i,
\end{equation}
%
%
\begin{equation}
{\partial}_{t}\overline{\rho}-\overline{u}_z+\overline{u}_{j}{\partial}_{j}\overline{\rho}=\frac{1}{Pr_0}\frac{1}{Re_{0}}{\partial}_{jj}\overline{\rho}-{\partial}_{j}\lambda_{j}
\end{equation}
$\overline{p}$ is given by $\overline{p} = \overline{p_f}$ +$\frac{1}{3}\overline{\rho}\tau_{ii}^R$, where $p_f$ is the filtered pressure and $\tau_{ij}^R = \overline{u_i u_j}-\overline{u}_i \overline{u}_j$.
$Re_0$, $Fr_0$, $Pr_0$, $Ro_0$ are the reference Reynolds number($= \frac{\mathcal{U L}}{\nu}$), Froude number($ = \frac{\mathcal{U}}{\mathcal{LN}}$ ), Prandtl number($= \frac{\nu}{\alpha}$) and Rossby Number($= \frac{\mathcal{U}}{2 \Omega \mathcal{ L}}$), respectively. Here, $\mathcal{U, L}$ are the characteristic velocity and length scales. $\mathcal{N}$ is the Brunt-Vaisala frequency($ = \sqrt{-\frac{g}{\rho_o} \frac{\partial \rho (z)}{\partial z}}$), and $\Omega$ is the angular rotation speed along the $z$-axis. $\nu$ and $\alpha$ are the kinematic viscosity and thermal diffusivity, respectively. $g$ is gravitational acceleration and $\rho_0$ is the density at ``reference" conditions. Details regarding the SGS stress, $\tau_{ij}$, SGS density flux $\lambda_j$ and SGS models such as the constant coefficient, the dynamic Smagorinsky and the dynamic Clark models \cite{GerPioMoiCab91,lilly1992proposed, wang2005dynamic} are described next.


The form of the filtered equations is dependent on the SGS model. The eddy viscosity-based models are written as follows:
 \begin{equation}
 \tau_{ij}= -2\nu_{T}\overline{S}_{ij} \hspace{0.5cm} 
 \overline{S}_{ij}=\frac{1}{2}(\frac{\partial{\overline{u}_{i}}}{\partial{x_{j}}}+\frac{\partial{\overline{u}_{j}}}{\partial{x_{i}}})
 \hspace{0.5cm}
 \nu_{T}=(C_{s}\Delta)^2\mid{\overline{S}}\mid{}
 \end{equation}
where, $\overline{S}_{ij}$ is the filtered rate of strain, $\nu_{T}$ is the eddy viscosity coefficient, $C_s$ is the Smagorinsky coefficient, $\mid \overline{S} \mid =(2\overline{S_{ij}}\;\overline{S_{ij}})^{1/2}$ and the SGS density flux, $\lambda_{j}$, is given by:
 \begin{equation}
 \lambda_{i}=-\alpha_{T}{\partial}_{i}\overline{\rho}\hspace{1cm}\alpha_{T}=(C_{\rho}\Delta)^2\mid \overline{S} \mid  \hspace{1cm} C^2_{\rho} = \frac{C^2_{s}}{Pr_T}   
 \end{equation}
where, $\alpha_{T}$ is constant and $Pr_{T}$ is turbulent Prandtl number. For the constant coefficient Smagorinsky model, the value of $C_{s}=0.16$ which was tuned for homogeneous isotropic turbulence \cite{pope2001turbulent,LILLY67} has been used in this work. Analogous to $C_s$, $C_{\rho}$ is also another constant as defined above. The value of $Pr_{T}$ is well established as $Pr_T \approx 0.7$ \cite{LILLY67} for neutral stratification. 

The dynamic Smagorinsky model evaluates the model coefficients, $C_s, \; C_{\rho}$ using information available to the flow by relying upon the Germano identity \cite{GerPioMoiCab91},
 \begin{equation}
 L_{ij}=T_{ij}-\widehat{\tau}_{ij}=\widehat{\overline{u}_{i}\overline{u}_{j}}-\hat{\overline{u}}_{i}\hat{\overline{u}}_{j}
 \end{equation}
where, $L_{ij}$ is the resolved stress tensor, $T_{ij}$ is the residual stress based on double-filtering: ($\widehat{...}$) represents a test-filtered field with filter width of $\widehat{\Delta}=2\Delta$ (in this work, a sharp-spectral filter is employed). Based on these equations $C_{s}$ and $C_{\rho}$ can be dynamically evaluated as \cite{GerPioMoiCab91,lilly1992proposed}\
 \begin{equation} \label{eq:Cs}
 C^2_{s}=\frac{\langle{L_{ij}M_{ij}}\rangle}{2 \langle{ M_{ij}M_{ij}}\rangle} 
 \end{equation}
where, $M_{ij}=\widehat{\Delta}^2\widehat{\mid{\overline{S}}\mid\overline{S_{ij}}}-\Delta^2\widehat{\mid{}\overline{S}\mid}\widehat{\overline{S}}_{ij}$
 and
 \begin{equation}
 C^2_{\rho}=\frac{\langle{\varepsilon_{ij}Q_{ij}}\rangle}{2\langle{Q_{ij}Q_{ij}}\rangle} \hspace{0.8cm}
Q_{i}=\widehat{\Delta}^2 \; \widehat{|\overline{S}| \; \partial_{i}\overline{\rho}}
- \Delta^2\widehat{\mid{}\overline{S}\mid}\widehat{{\partial}_{i}\overline{\rho}}
 \end{equation}
 \begin{equation}
 \varepsilon_{i}=L_{ii}-\widehat{\lambda}_{i}=\widehat{\overline{u}_{i}\overline{\rho}}-\widehat{\overline{u}}_{i}\widehat{\overline{\rho}}
 \end{equation}

The third model considered in this work is the dynamic, mixed, non-linear model proposed by Leonard \cite{LEONARD1975237}, Clark et al. \cite{CRF79} and Liu et al. \cite{LSM95}. The model formulation is given as, 
\begin{equation}
    \tau_{ij}^{nl}=C_{nl}\Delta^{2} \overline{A_{ki}} \overline{A_{kj}}
\end{equation}
where, $\overline{A_{ki}}=\frac{\partial \overline{u_{i}}}{\partial x_{k}}$, $C_{nl}$ is the model coefficient and is a function of filter type and test filter scale. This model is able to exhibit a limited amount of backscatter of energy. For computational reasons, in practice, an additional eddy viscosity is added to stabilize the solutions. This also aids in improving the geometric alignment of the SGS stress eigenvectors (Katz et al. \cite{KATZ02}). In this form, the mixed non-linear model is given by \cite{AMC99}:
\begin{equation}
    \tau_{ij}^{mnl}=\tau_{ij}^{smg}+\tau_{ij}^{nl}=-2(C_{s}\Delta)^{2}\mid{\overline{S}}\mid{}\overline{S}_{ij}+C_{nl}\Delta^{2} \overline{A_{ki}} \overline{A_{kj}}
\end{equation}
The dynamic coefficients $C_{s}$ and $C_{nl}$ are estimated by modifying resolved stresses as follows:
\begin{equation}
    L_{ij}=C_{s}^{2} M_{ij} + C_{nl} N_{ij}
\end{equation}
where,
\begin{equation}
    N_{ij}=\widehat{\Delta}\widehat{\overline{A_{ki}}}\widehat{\overline{A_{kj}}}-\Delta^{2}\widehat{\overline{A_{ki}} \overline{A_{kj}}}
\end{equation}
Similar to Equation \ref{eq:Cs}, using $L_2$ norm minimization of the error function leads to, 
\begin{equation}
    C_{s}^{2}=\frac{\langle{L_{ij}M_{ij}}\rangle\langle{N_{ij}N_{ij}}\rangle-\langle{L_{ij}N_{ij}}\rangle\langle{M_{ij}N_{ij}}\rangle}{\langle{M_{ij}M_{ij}}\rangle\langle{N_{ij}N_{ij}}\rangle-\langle{M_{ij}N_{ij}}\rangle^{2}}
    \label{Cs}
\end{equation}
\begin{equation}
    C_{nl}=\frac{\langle{L_{ij}N_{ij}}\rangle\langle{M_{ij}M_{ij}}\rangle-\langle{L_{ij}M_{ij}}\rangle\langle{M_{ij}N_{ij}}\rangle}{\langle{M_{ij}M_{ij}}\rangle\langle{N_{ij}N_{ij}}\rangle-\langle{M_{ij}N_{ij}}\rangle^{2}}
    \label{Cn}
\end{equation}

\section{Problem Description}

The problem considered here is an initially-isotropic decaying turbulence case. Calculations are initialized using a divergence-free velocity field with an energy spectrum given by:
\begin{equation}
E(\kappa) = C_k\kappa^{4} exp[-\frac{1}{2} \left( \frac{\kappa}{\kappa_0}^2 \right)],
\end{equation}
where $\kappa$ is the wavenumber magnitude, the value of the constant, $C_k$ is chosen such that the total turbulent kinetic energy associated with the initial velocity field is 0.5. The spectrum is centered on $\kappa_0$ with $\kappa_0 \in [2,3]$. There is no initial potential energy, but it develops due to energy exchange. The velocity field is initialized using random phase Fourier modes but scaled in Fourier space so that the energy spectrum corresponds to the initial energy spectrum. From this initially random isotropic field we perform an isotropic precalculation and so provide a realistic velocity field as the starting point of the anisotropic runs.

\section{Computational details}
A $(2 \pi)^3$ triply periodic, incompressible, pseudo-spectral code is used here \cite{doi:10.1080/14685248.2015.1078469,agrawal2021large,jadhav2021assessment}. Cylindrical truncation is performed using the standard $\frac{2}{3}$-dealiasing rule. Third-order explicit Runge-Kutta time stepping scheme with a CFL criterion is used for the temporal integration of the governing equations in spectral space. 
The time step, $\Delta{t}$, is defined as, $\Delta{t}=\frac{cfl\ 2 \pi}{N \sqrt{tke_{0}}}$, and is kept constant throughout the simulation, where $tke_{0}$ is the initial turbulent kinetic energy, and $cfl$ is chosen to be 0.05 for all the cases \cite{pope2001turbulent}. The low value of the $cfl$ is related to the fact that the time step is defined using only the stratification time scale, and not the other relevant scales in the flow.

\section{Results}

LES calculations of decaying stratified-rotating turbulence using different SGS models are presented here in this section. This section begins with the LES studies for grid dependence, followed by comparisons with in-house DNS results.

\subsection{Grid study}
In order to assess the grid dependence of LES, the dynamic Smagorinsky-based LES calculations at three different grid densities are compared with in-house DNS calculations. The parameters of the DNS are as follows: the Brunt-Vaisala frequency ($\mathcal{N}$) is 5.5, the Coriolis rotation frequency ($f$) of 0.04, and the Reynolds number ($Re_0$) is 3704. The DNS calculations used a grid of size $768^{3}$ ($\kappa_max \eta = 1.45$), whereas the grids employed for LES are $48^{3}$, $96^{3}$ and $192^{3}$.  For consistent comparisons of certain quantities, such as turbulent kinetic energy, between computations at various resolutions, all the SGS models are considered in conjunction with a spectral cutoff filter of a specific filter width.  Hence, the corresponding filter widths for grids of $48^{3}$, $96^{3}$ and $192^{3}$ points are $\pi/16$, $\pi/32$ and $\pi/64$, respectively.  The effect of grid refinement on the results of LES models is studied by analyzing turbulent kinetic energy ($tke$) and kinetic dissipation ($\epsilon_k$). The $tke$ is calculated as,
\begin{equation}
{tke(t)=\int_{0}^{\infty}E_{k}(\kappa,t)d\kappa=\frac{1}{2}\sum_{\kappa}\widehat{u}(\kappa,t)\widehat{u}^{*}(\kappa,t)}
\end{equation}
Also, $\epsilon_{k}$ is the dissipation of $tke$ and is given by:
\begin{equation}
{\epsilon_{k}(t)=2\nu\int_{0}^{\infty}\kappa^{2}E_{k}(\kappa,t)d\kappa=\nu\sum_{\kappa}\kappa^{2}\widehat{u}(\kappa,t)\widehat{u}^{*}(\kappa,t)},
\end{equation}
where $\kappa=\mid \mathbf{\kappa} \mid$ is the wavenumber magnitude and $\mathbf{\kappa}=\kappa_{x}\hat{i}+\kappa_{y}\hat{k}+\kappa_{z}\hat{k}$ is the wavenumber vector.  $(\widehat{...})$ denotes a Fourier transform and $(...)^{*}$ denotes its complex conjugate and summation is taken over all wavenumbers. 

\begin{figure*}[!ht]
\begin{center}
\subfigure[]{
\label{comptke}
\includegraphics[scale=0.3]{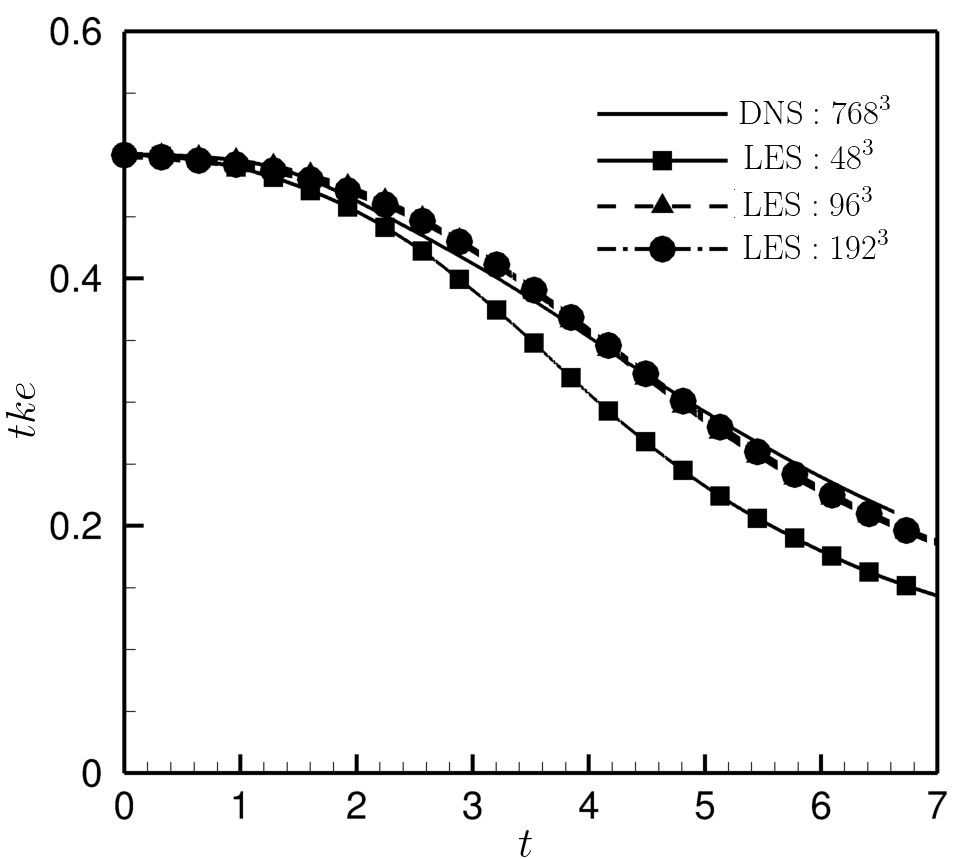}}
\subfigure[]{
\label{compek}
\includegraphics[scale=0.3]{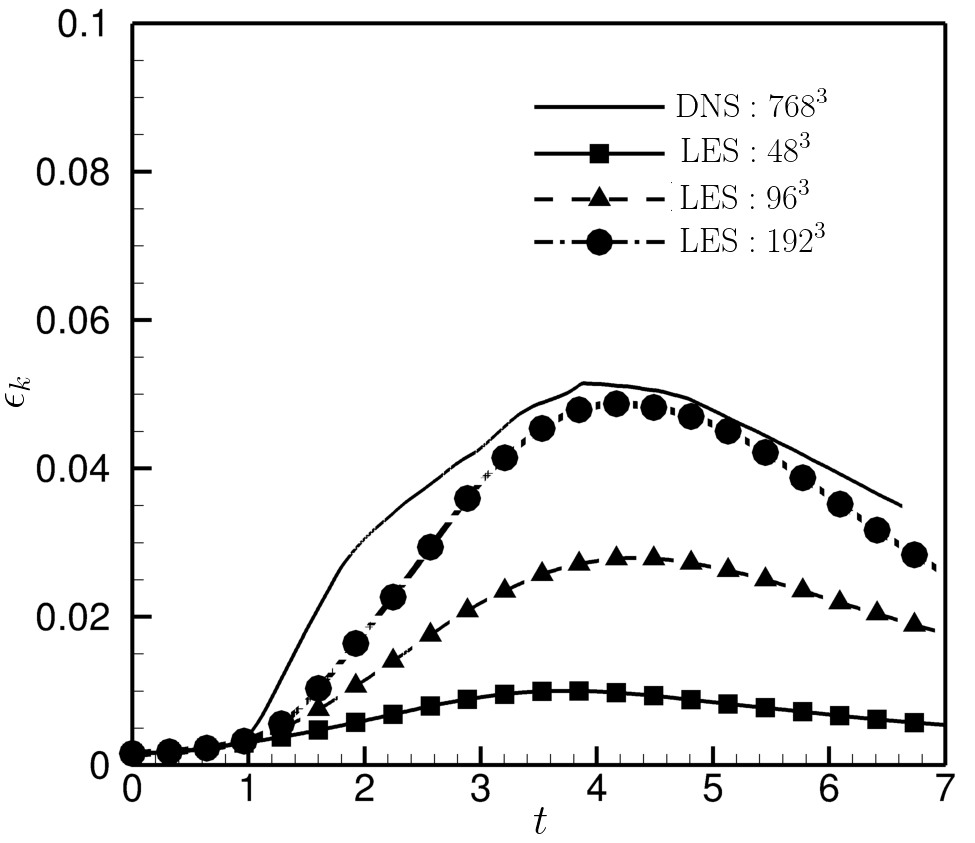}}
\subfigure[]{
\label{compekspec}
\includegraphics[scale=0.3]{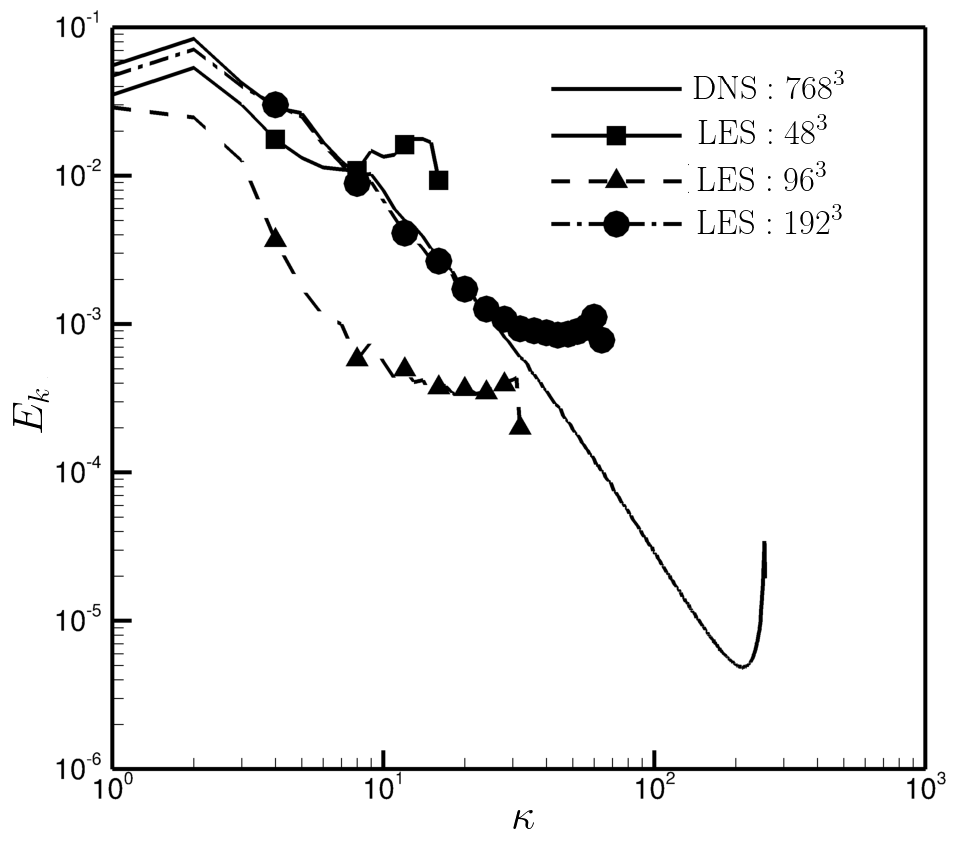}}
\caption{Evolution of (a) $tke$ and (b) $\epsilon_k$, and (c) kinetic energy spectra using the dynamic Smagorinsky model for grid densities, $48^3$ (solid), $96^3$ (dashed), and $192^3$ (dotted). The reference (or DNS) values are shown in black.}
\label{compare}
\end{center}
\end{figure*}

Figures \ref{comptke} and \ref{compek} compare the turbulent kinetic energy, $tke$, and its dissipation rate, $\epsilon_k$, respectively, between filtered DNS and LES calculations (at the three different grid densities). The trend followed by all grid densities is qualitatively similar to the DNS results. As the grid is refined, the errors decrease as expected. Figure \ref{compekspec} shows the kinetic energy spectra at the point of maximum dissipation.  The spectra plot (in Figure \ref{compekspec}) shows a pile-up of energy at $\kappa \approx 10$ for $48^{3}$, which thereby shifts to higher wavenumbers for $96^3$ and further for $192^{3}$. Also, the large-scale behavior is reasonably well predicted with $192^3$ degrees of freedom. Based on these results, a grid density of $192^3$ is chosen as the grid for all the LES calculations presented from here on. 


\subsection{Performance of LES models}

LES calculations of the stratified-rotating case using the constant coefficient Smagorinsky, the dynamic Smagorinsky, and the modified dynamic non-linear gradient-based (Clark-type) LES models are presented in this section. Some details of the reference cases are given in Table \ref{tab:param}. 

\begin{table}[!ht]
\centering
\begin{tabular}{|c|c|c|c|c|c|c|}
\hline
Case & $\mathcal{N}$($s^{-1}$) & f($s^{-1}$) & $\mathcal{N}$/f & $Ro_0$ & $Fr_0$ & $Re_0$\\ \hline
1 & 0.2 & 0.04 & 5 & 25 & 5 & 3704\\
2 & 1.6 & 0.04 & 40 & 25 & 0.625 & 6667 \\
3 & 5.5 & 0.04 & 138 & 25  & 0.1818 & 6667\\ \hline \hline
\end{tabular}
\caption{A brief description of the relevant dimensional, and non-dimensional quantities which are simulated in this work. }
\label{tab:param}
\end{table}

The cases are chosen such that $\mathcal{N}/f$ varies from $\mathcal{O}(1)$ to $\mathcal{O}(100)$, in order to analyze the significant changes in the effects of stratification relative to rotation. A lower ratio has not been studied, as the authors believe this would primarily result in a rotation-driven regime, which has been studied previously \cite{KOB01,LU07}. 

\begin{figure*}[!ht]
\begin{center}
\subfigure[]{
\label{tke0402}
\includegraphics[scale=0.34]{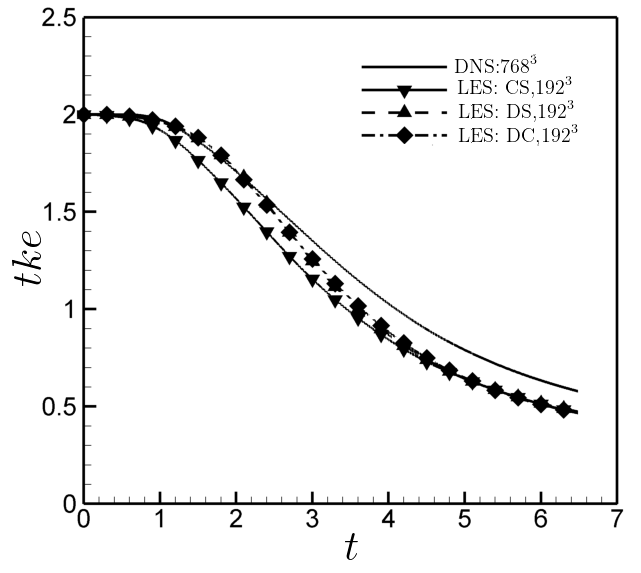}}
\subfigure[]{
\label{tke0416}
\includegraphics[scale=0.34]{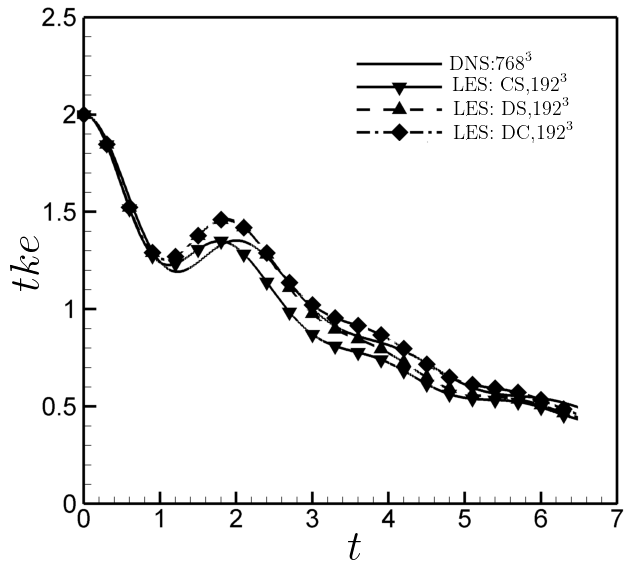}}
\subfigure[]{
\label{tke0455}
\includegraphics[scale=0.34]{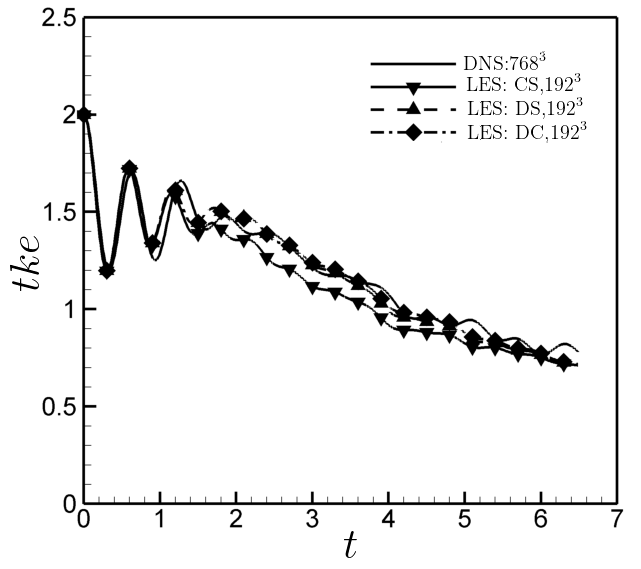}}
\subfigure[]{
\label{tpe0402}
\includegraphics[scale=0.34]{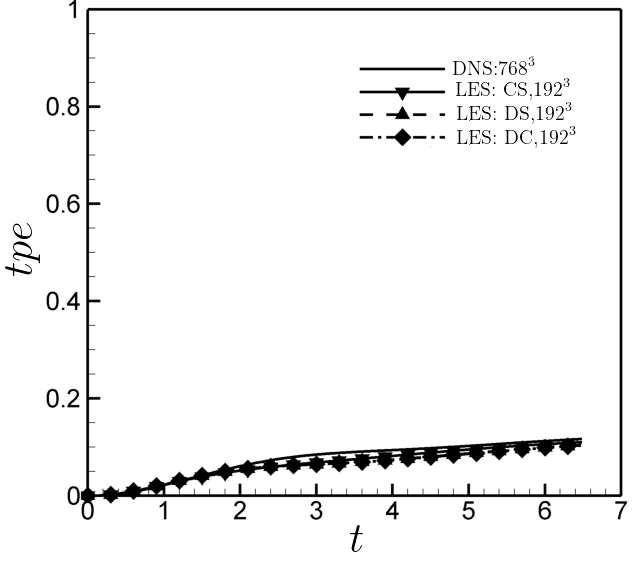}}
\subfigure[]{
\label{tpe0416}
\includegraphics[scale=0.34]{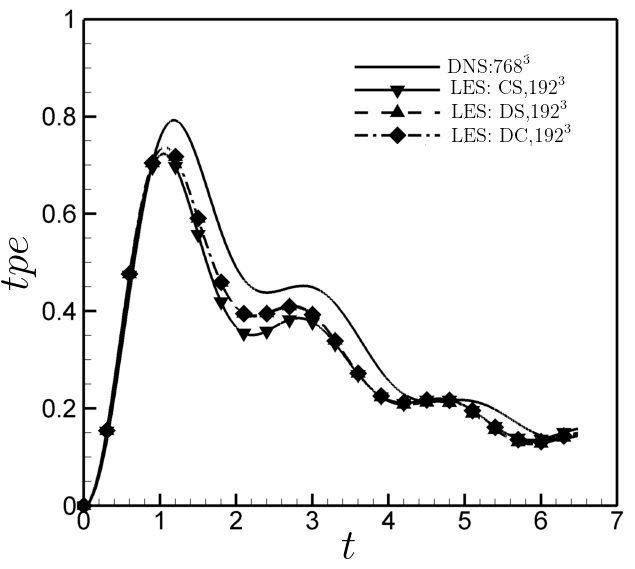}}
\subfigure[]{
\label{tpe0455}
\includegraphics[scale=0.34]{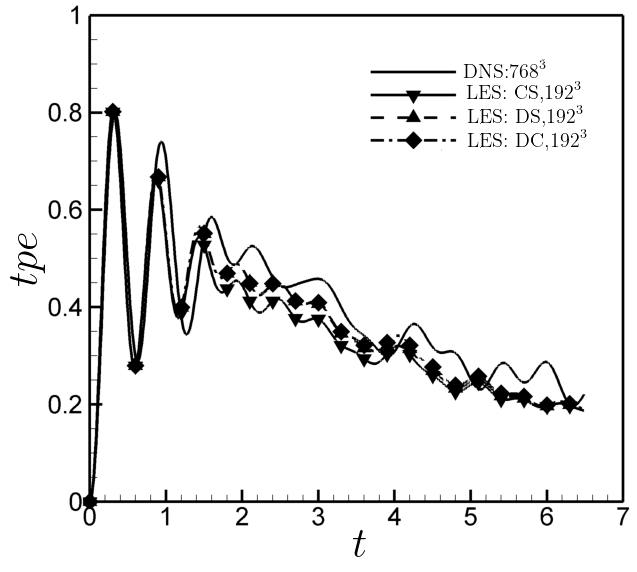}}
\caption{Evolution of $tke$ using constant coefficient Smagorinsky (CS), dynamic Smagorinsky (DS), and dynamic Clark (DC) models for (a) Case 1, (b) Case 2, and (c) Case 3. Evolution of $tpe$ using constant coefficient Smagorinsky (CS), dynamic Smagorinsky (DS), and dynamic Clark (DC) models for (d) Case 1, (e) Case 2, and (f) Case 3.}
\label{te04}
\end{center}
\end{figure*}

Figure \ref{te04} shows the evolution of turbulent kinetic energy ($tke$) and turbulent potential energy ($tpe$) for all three flow cases considered. Mathematically, $tpe$ is defined as,
\begin{equation}
{tpe(t)=\int_{0}^{\infty}E_{p}(\kappa,t)d\kappa=\frac{1}{2}\sum_{\kappa}\widehat{\rho}(\kappa,t)\widehat{\rho}^{*}(\kappa,t)}
\end{equation}
The trend followed by LES models is qualitatively similar to that of the DNS for all values of $\mathcal{N}/f$. The mixed Clark model performs better than the Smagorinsky-type models in comparison, as observed in the $tke$ plot, while the $tpe$ is under-predicted by all LES models. As expected, the constant coefficient Smagorinsky model largely differs from the filtered DNS. Long-time scale oscillations are observed in the evolution of $tke$ and $tpe$ as the ratio $\mathcal{N}/f$ increases from left to right, with no fluctuations observed when the effect of rotation and stratification are comparable (Case 1). These fluctuations are observed due to the exchange of energy between kinetic and potential counterparts, dominant at higher $\mathcal{N}/f$. This is further confirmed by the fact that the two quantities are out of phase. The exchange of energy is a large-scale phenomenon; up to $t\approx 1$, the large scales don't break down and the LES results coincide with the DNS results. But for $t>1$, the deviation of LES results from DNS is observed. Also, the exchange between $tke$ and $tpe$ shows oscillations for Case 3 DNS results, but they are absent for LES results ($t>3$).   

\begin{figure*}[!ht]
\begin{center}
\subfigure[]{
\label{eps0402}
\includegraphics[scale=0.34]{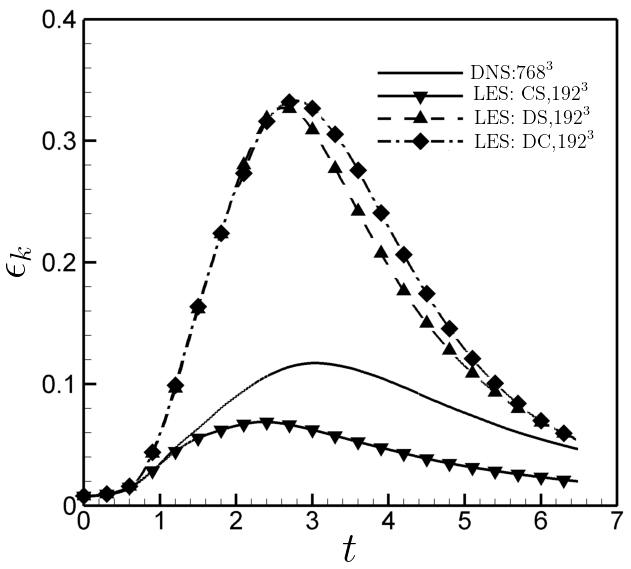}}
\subfigure[]{
\label{eps0416}
\includegraphics[scale=0.34]{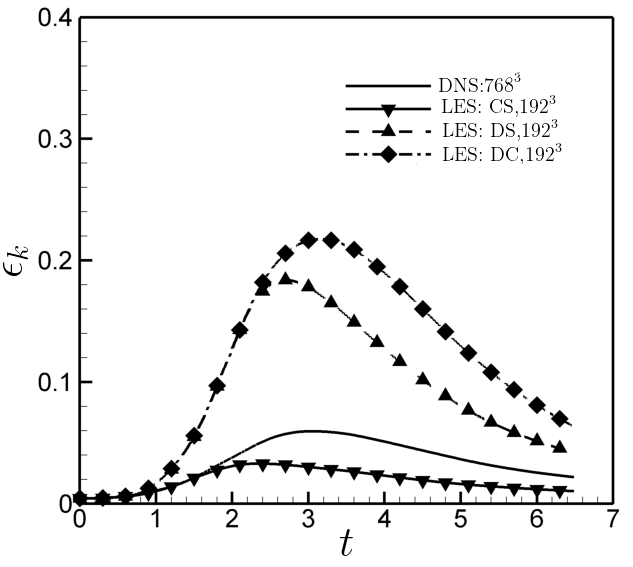}}
\subfigure[]{
\label{eps0455}
\includegraphics[scale=0.34]{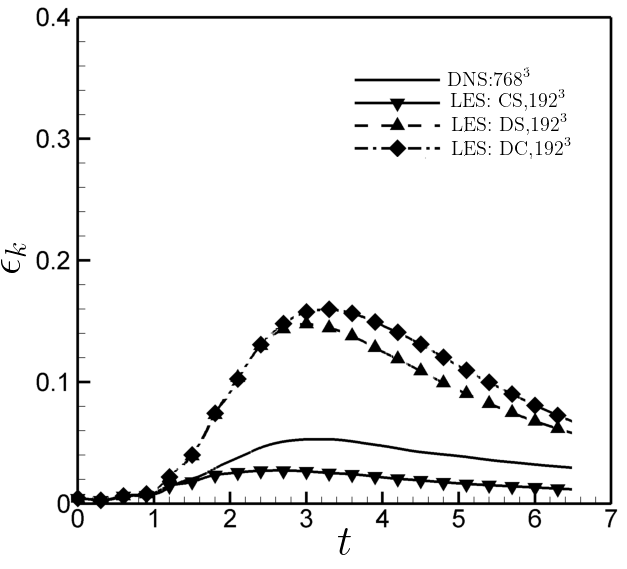}}
\caption{Evolution of kinetic energy dissipation using constant coefficient Smagorinsky (CS), dynamic Smagorinsky (DS), and dynamic Clark (DC) models for (a) Case 1 (b) Case 2 (c) Case 3.}
\label{epsk04}
\end{center}
\end{figure*}

\begin{figure*}[!ht]
\begin{center}
\subfigure[]{
\label{epsp0402}
\includegraphics[scale=0.34]{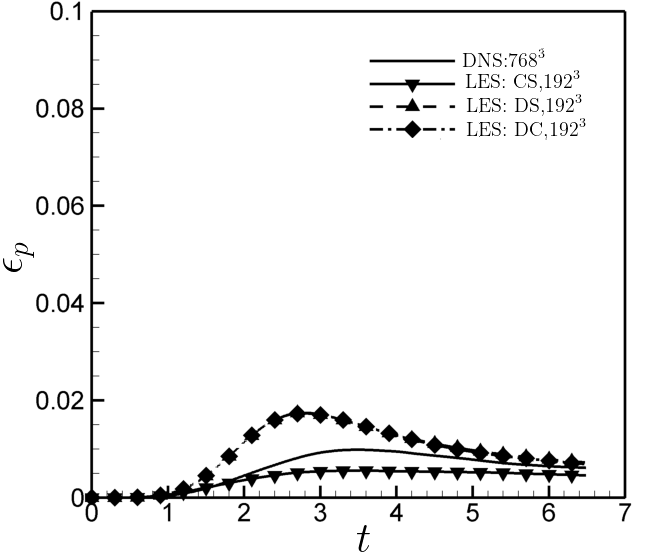}}
\subfigure[]{
\label{epsp0416}
\includegraphics[scale=0.34]{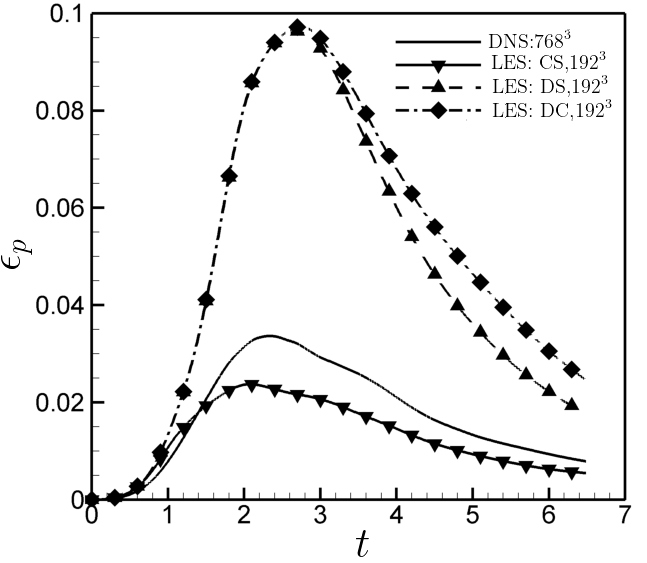}}
\subfigure[]{
\label{epsp0455}
\includegraphics[scale=0.34]{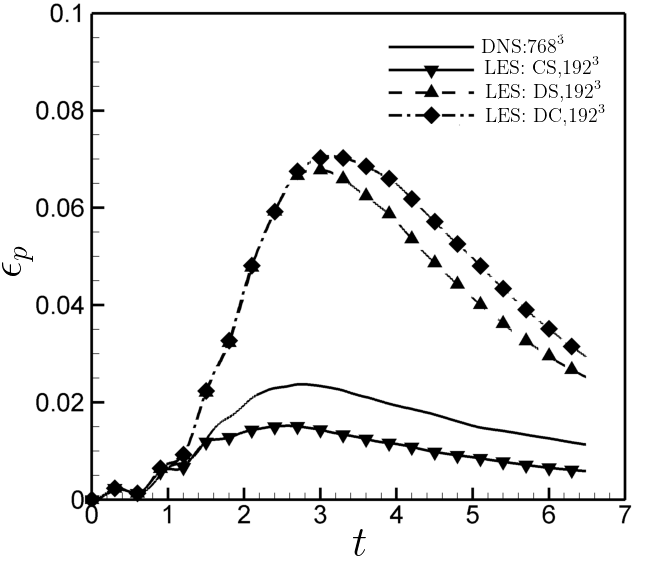}}
\caption{Evolution of potential energy dissipation using constant coefficient Smagorinsky (CS), dynamic Smagorinsky (DS), and dynamic Clark (DC) models for (a) Case 1 (b) Case 2 (c) Case 3.}
\label{epsp04}
\end{center}
\end{figure*}

The evolution of kinetic and potential dissipation is shown in Figures \ref{epsk04} and \ref{epsp04}. The potential dissipation is given by,
\begin{equation}
{\epsilon_{p}(t)=2\frac{\nu}{Pr_0}\int_{0}^{\infty}\kappa^{2}E_{p}(\kappa,t)d\kappa=\frac{\nu}{Pr_0} \sum_{\kappa}\kappa^{2} \widehat{\rho}(\kappa,t)\widehat{\rho}^{*}(\kappa,t),}
\end{equation}

For all the simulations, the value of the kinetic energy dissipation ($\epsilon$) increases from a small positive quantity to a maximum and then decreases. This is very well in line with the expectations as the flow is initially laminar, and it transitions to turbulence before eventually decaying. The maximum value reached for kinetic dissipation decreases as the value of $\mathcal{N}/f$ value increases. Presumably, this is related to the inhibited turbulent scales in the vertical direction due to the increased stratification. The time instance at which the maximum value occurs is at $t=3$ for all cases. LES employing either of the two dynamic subgrid-scale models follows the same trend as DNS, however, for the simulations using the constant coefficient Smagorinsky model, the predictive trends are poor in both the maximum value, and the time at which the maximum value is attained. The dynamic Smagorinsky and dynamic Clark model dissipation values are $\approx 3 \times$ the DNS values. Thus, as the simulation evolves, a large amount of energy is transferred towards small scales (explained further using the distribution of spectra later). The kinetic energy dissipation is the lowest for Case 3, and similar observations are made for potential energy dissipation.

One of the major characteristics of stratified-rotating flows involves the differences in horizontal and vertical length scales. The integral horizontal and vertical scales are defined as \cite{lindborg_2006}:
\begin{equation}
l_{h}=\frac{2\pi\int_{0}^{\infty}E(\kappa_{h})d\kappa_{h}}{\int_{0}^{\infty}\kappa_{h}E(\kappa_{h})d\kappa_{h}}
\end{equation}
\begin{equation}
l_{v}=\frac{2\pi\int_{0}^{\infty}E(\kappa_{v})d\kappa_{v}}{\int_{0}^{\infty}\kappa_{v}E(\kappa_{v})d\kappa_{v}}
\end{equation}

\begin{figure*}[!ht]
\begin{center}
\subfigure[]{
\label{lh0402}
\includegraphics[scale=0.34]{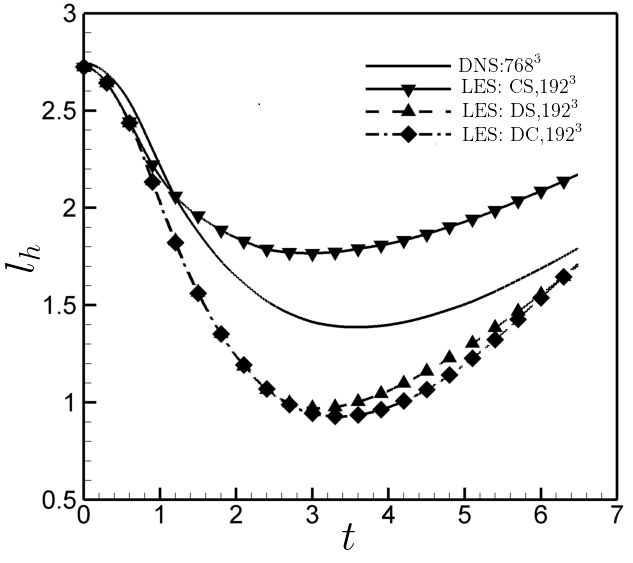}}
\subfigure[]{
\label{lh0416}
\includegraphics[scale=0.34]{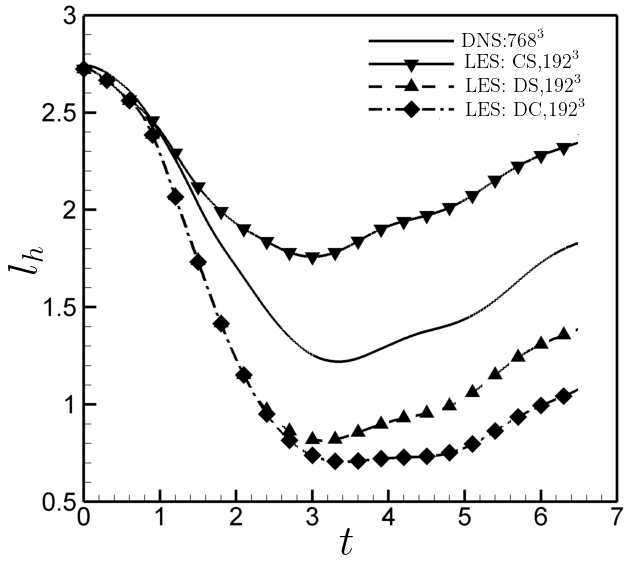}}
\subfigure[]{
\label{lh0455}
\includegraphics[scale=0.34]{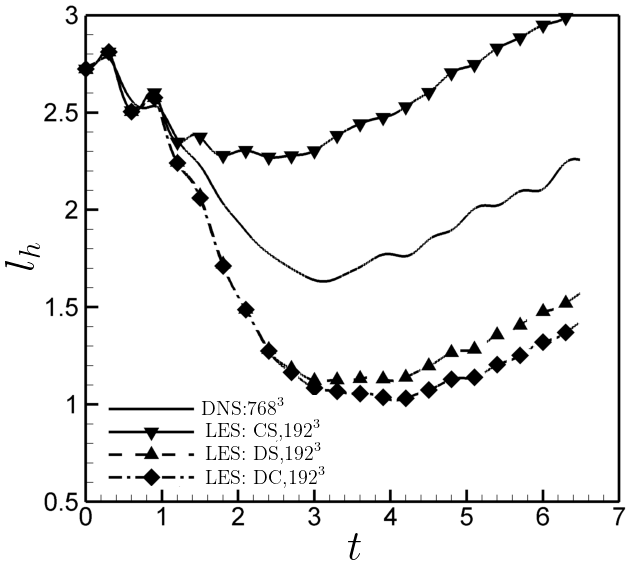}}
\caption{Evolution of horizontal length scale using constant coefficient Smagorinsky (CS), dynamic Smagorinsky (DS), and dynamic Clark (DC) models for (a) Case 1 (b) Case 2 (c) Case 3. The reference values are shown in black.}
\label{lh04}
\end{center}
\end{figure*}

\begin{figure*}[!ht]
\begin{center}
\subfigure[]{
\label{lv0402}
\includegraphics[scale=0.34]{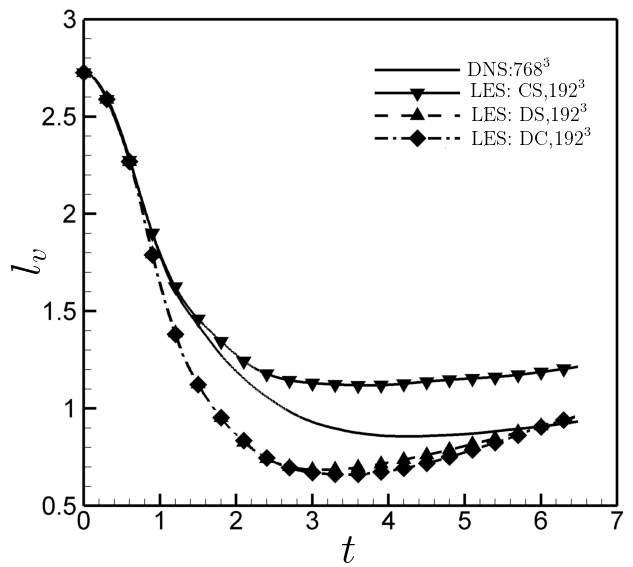}}
\subfigure[]{
\label{lv0416}
\includegraphics[scale=0.34]{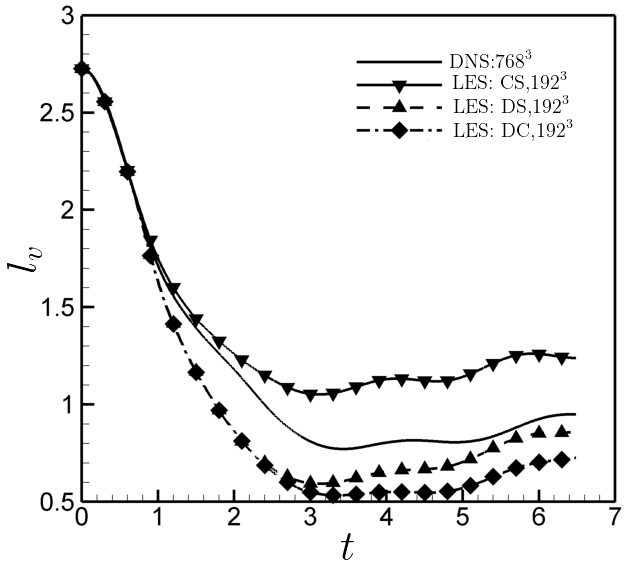}}
\subfigure[]{
\label{lv0455}
\includegraphics[scale=0.34]{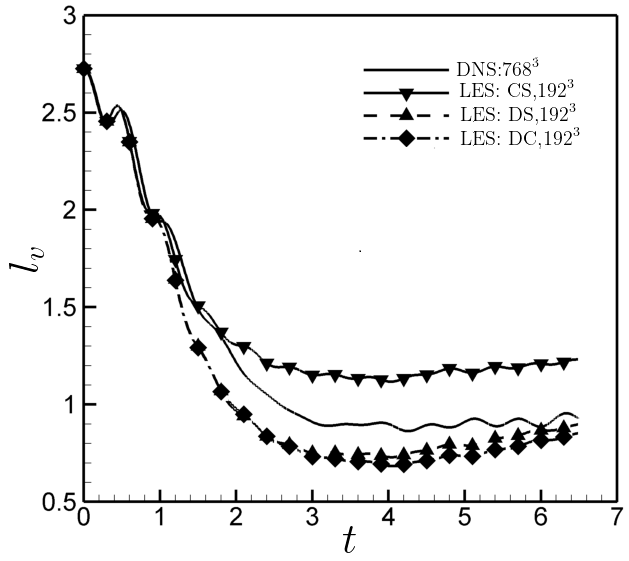}}
\caption{Evolution of vertical length scale using constant coefficient Smagorinsky (CS), dynamic Smagorinsky (DS), and dynamic Clark (DC) models for (a) Case 1 (b) Case 2 (c) Case 3. The reference values are shown in black.}
\label{lv04}
\end{center}
\end{figure*}

Figures \ref{lh04} and \ref{lv04} show the evolution of $l_{h}$ and $l_{v}$, respectively. It can be seen that the horizontal and vertical length scales have initially large values. The length scales decrease when the large scales break down into smaller scales as the flow transitions to turbulence. The lowest values of $l_{h}$ and $l_{v}$ occur approximately at $t \approx 4.5$ for Case 1 and Case 2, which coincide with the point of maximum energy dissipation.  The horizontal length scale is larger for Case 3, due to an increase in stratification; this forces the smaller scales to contain less energy and thus shifts the integral measure of the length scales to a lower wavenumber. After the minimum is reached, an increase in length scales is observed. The value of the horizontal length scale predicted by the two dynamic LES models is lower than filtered DNS, whereas the constant-coefficient model overpredicts the length scales. This indicates that the breakdown of the large scales is minimal for the classical Smagorinsky model, and this is also reflected in the lower dissipation values in Figure \ref{epsp04}. The trend and values of $l_v$ are captured correctly by dynamic Smagorinsky and dynamic Clark models, with the minimum value being similar for all three simulations.

\begin{figure*}[!ht]
\begin{center}
\subfigure[]{
\label{Ek0402}
\includegraphics[scale=0.34]{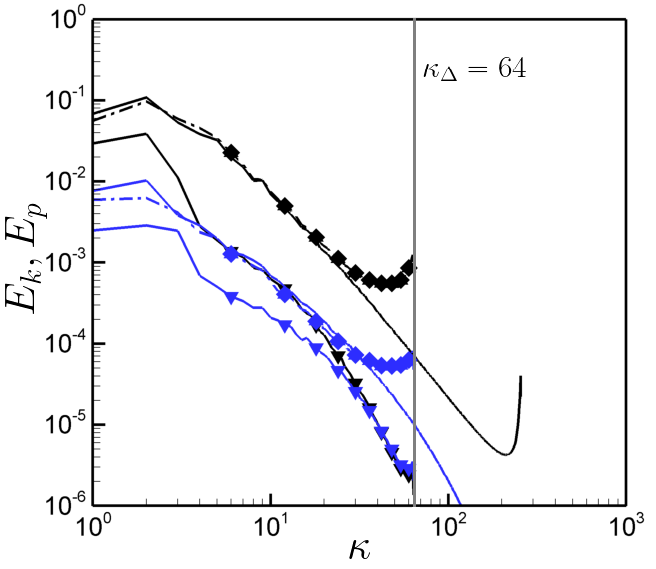}}
\subfigure[]{
\label{Ek0416}
\includegraphics[scale=0.34]{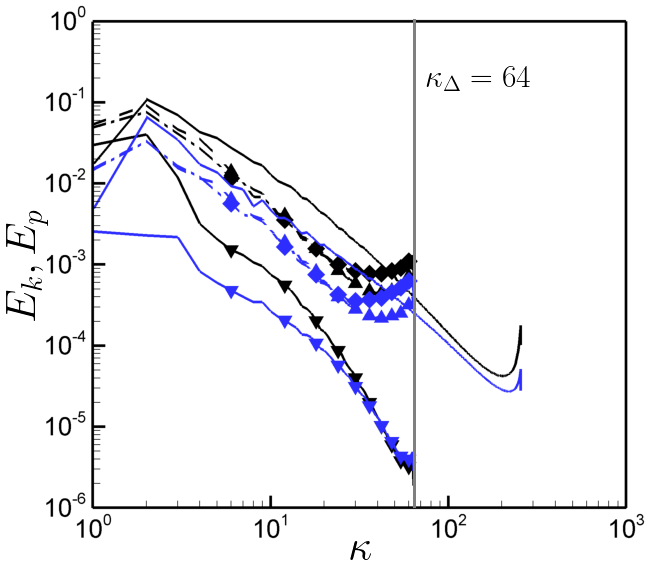}}
\subfigure[]{
\label{EK0455}
\includegraphics[scale=0.34]{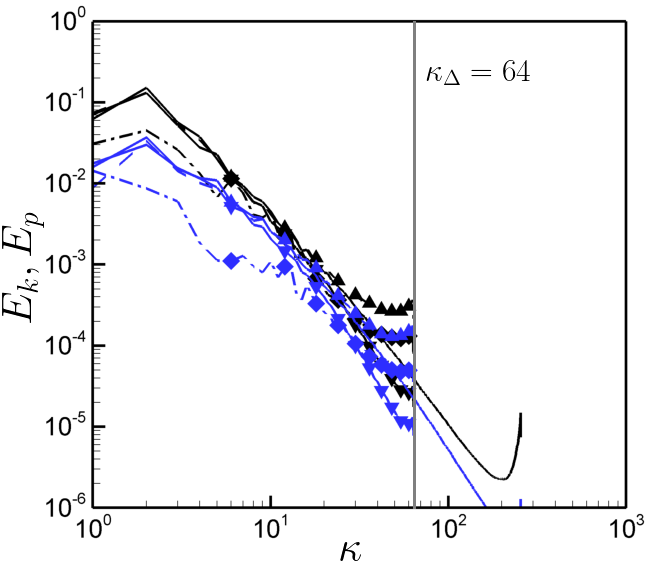}}
\caption{Kinetic and potential energy spectra using constant coefficient Smagorinsky (CS), dynamic Smagorinsky (DS), and dynamic Clark (DC) models at the point of maximum dissipation for (a) Case 1 (b) Case 2 (c) Case 3. The legends are the same as in figure \ref{te04}. $E_k$ (Black), $E_p$ (Blue).}
\label{Ek04}
\end{center}
\end{figure*}

Figure \ref{Ek04} shows the kinetic energy spectra, $E_k(\kappa)$, and the potential energy spectra, $E_p(\kappa)$, respectively, for the filtered DNS and LES calculations at the point of maximum energy dissipation. Both the dynamic SGS models reasonably predict the kinetic energy spectra for $\kappa < 25$ for Case 1. For Cases 2 and 3, even the two dynamic models misrepresent the kinetic energy spectra for $\kappa > 10$. The constant coefficient Smagorinsky model poorly predicts the energy spectra for all cases. The lower value of dissipation for the constant coefficient Smagorinsky model can be explained by lower energy contained at higher wavenumbers ($\kappa>20$). Similar observations can be made for the potential energy spectra.  

\begin{figure*}[!ht]
\begin{center}
\subfigure[]{
\label{Fh0402}
\includegraphics[scale=0.49]{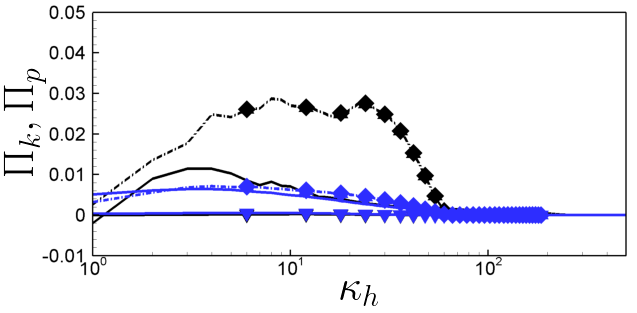}}
\subfigure[]{
\label{Fh0416}
\includegraphics[scale=0.49]{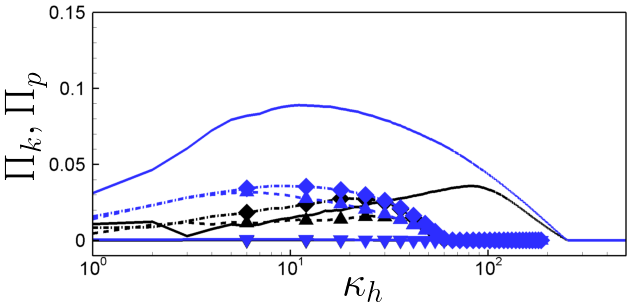}}
\subfigure[]{
\label{Fh0455}
\includegraphics[scale=0.49]{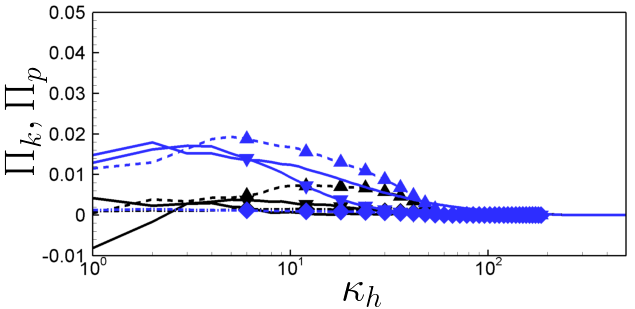}}
\caption{Energy transfer spectra in the horizontal direction using constant coefficient Smagorinsky (CS), dynamic Smagorinsky (DS), and dynamic Clark (DC) models at the point of maximum dissipation (a) Case 1, (b) Case 2, and (c) Case 3. The legends are the same as in figure \ref{te04}. $\Pi_k(\kappa_h)$ (Black), $\Pi_p(\kappa_h)$ (Blue). }
\label{Fh04}
\end{center}
\end{figure*}

\begin{figure*}[!ht]
\begin{center}
\subfigure[]{
\label{Fv0402}
\includegraphics[scale=0.49]{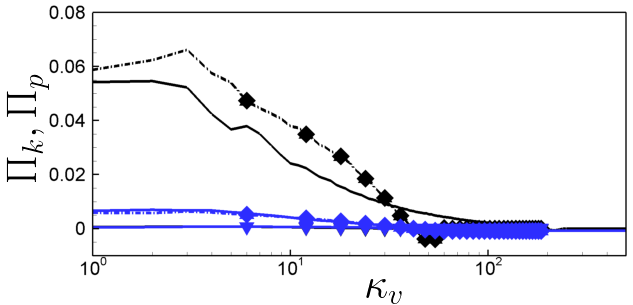}}
\subfigure[]{
\label{Fv0416}
\includegraphics[scale=0.49]{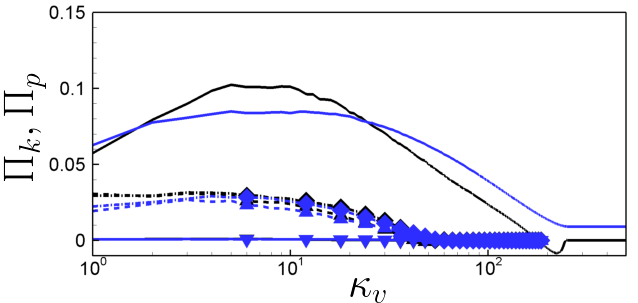}}
\subfigure[]{
\label{Fv0455}
\includegraphics[scale=0.49]{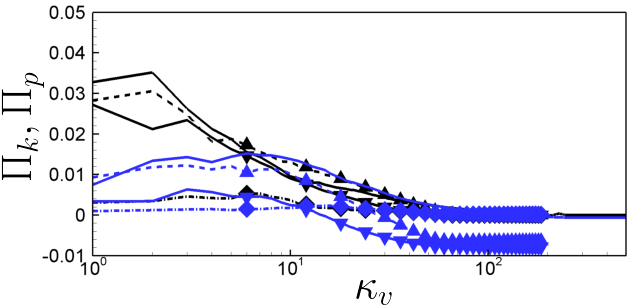}}
\caption{Energy transfer spectra in the vertical direction using constant coefficient Smagorinsky (CS), dynamic Smagorinsky (DS), and dynamic Clark (DC) models at the point of maximum dissipation (a) Case 1, (b) Case 2 and (c) Case 3. The legends are the same as in figure \ref{te04}. $\Pi_k(\kappa_v)$ (Black), $\Pi_p(\kappa_v)$ (Blue). }
\label{Fv04}
\end{center}
\end{figure*}

Energy transfer flux explains the energy cascade process in a flow. A positive value of the flux implies a forward cascade, i.e., net energy transfer from large scales to small scales, while a negative value points towards a reverse cascade. Figures \ref{Fh04} and \ref{Fv04} show the spectra of kinetic and potential energy flux with respect to vertical and horizontal wavenumbers. These quantities are defined based on the spectral energy transfer functions $\mathbf T_K(\mathbf{\kappa},t)$ and $ \mathbf T_P(\mathbf{\kappa},t)$ for kinetic and potential energies, respectively. For instance, the kinetic energy flux through the horizontal wavenumber, $\kappa_h$, is defined as,
\begin{equation}
\Pi_K(\kappa_h,t)=  - \sum_{\substack{
            \sqrt{\kappa_x^2+\kappa_y^2}{\leq}\kappa_h \\
            {\kappa_z}}  }
            \mathbf T_K(\mathbf{\kappa},t)
 \label{eqn:fluxh} 
\end{equation}
and kinetic energy flux through vertical wavenumbers, $\kappa_v$, as,
\begin{equation}
\Pi_K(\kappa_v,t)=  - \sum_{\substack{
            |\kappa_z|{\leq}\kappa_v \\
            {\kappa_x,\kappa_y}}  }
            \mathbf T_K(\mathbf{\kappa},t).
   \label{eqn:fluxv} 
\end{equation}
Similarly, the potential energy flux through the horizontal ($\Pi_P(\kappa_h,t)$) and vertical wavenumbers ($\Pi_P(\kappa_v,t)$) can be calculated from the spectral potential energy transfer function. The reader is referred to Lindborg \cite{lindborg_2006} for more details on the definitions of the spectral energy transfer functions.

Figure \ref{Fh04} shows that the kinetic energy flux across horizontal wavenumbers for LES differs significantly from the filtered DNS results at the point of maximum dissipation. For the filtered DNS, the range of wavenumbers over which the value of $Fkh$ is positive decreases as the $\mathcal{N}/f$ increases. For example, in Case 1 (Figure \ref{Fh0402}) the wavenumbers over which $Fkh \geq 0 $ is $2<\kappa<200$, while for Case 3 (Figure \ref{Fh0455}) the range reduces to $20<\kappa<200$. This is due to the domination of stratification on rotation as the $\mathcal{N}/f$ value increases, which causes an inverse energy cascade as seen from the negative values of $Fkh$ for Case 3 over the wavenumber range $2<\kappa<20$. On the contrary, the LES results show a positive value across the entire wavenumber range for all the cases. The results for the dynamic Smagorinsky and Clark models are close to each other. Even though the numerical value of $Fkh$ decreases as $\mathcal{N}/f$ increases, the negative fluxes for Cases 2 and 3 are not observed in contrast to the filtered DNS. Thus, a significant forward cascade is observed for the LES results for the dynamic Smagorinsky and Clark models. The forward cascade observed in LES indicates the transfer of energy from large scales to small scales, which eventually is dissipated (as seen in Figure \ref{epsk04}), and an overestimation of dissipation values is observed for LES results in comparison to DNS. The value of the dynamic coefficients (Equations \ref{Cs} and \ref{Cn}) is positive throughout the simulation, which leads to the dissipation of energy at the cutoff wavenumber ($\kappa_\Delta = 64$) and no reverse cascade of energy is predicted by the LES models. This further corroborates the observation that the horizontal length scale for LES cases is smaller than DNS. The potential energy flux across vertical wavenumbers is also plotted in Figure \ref{Fh04}. The LES and DNS values are comparable across the wavenumber range. The values of potential energy flux for dynamic Smagorinsky and Clark models agree with filtered DNS values across horizontal wavenumbers for all cases. The kinetic and potential flux across vertical wavenumbers is shown in Figure \ref{Fv04}. The values of DNS and LES are comparable in the wavenumber range $\kappa<30$ for Case 1. However, for both Cases 2 and 3, the flux values predicted from LES do not agree with the DNS. The above observations for the flux of kinetic and potential energy across horizontal and vertical wavenumbers indicate that as the stratification dominates over rotation from Case 1 to Case 3, the inverse cascade starts to dominate (as seen from the horizontal kinetic flux in the DNS results in Figure \ref{Fh04}). This is opposite to the behaviour exhibited by LES where a forward cascade is observed for all cases. 

\begin{figure*}[!ht]
\begin{center}
\subfigure[]{
\label{con0402}
\includegraphics[scale=0.295]{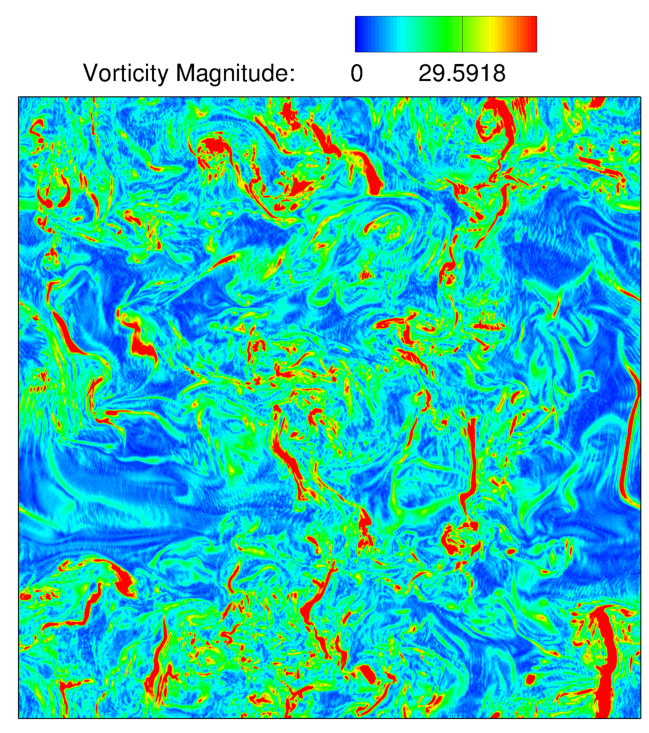}}
\subfigure[]{
\label{con0416}
\includegraphics[scale=0.295]{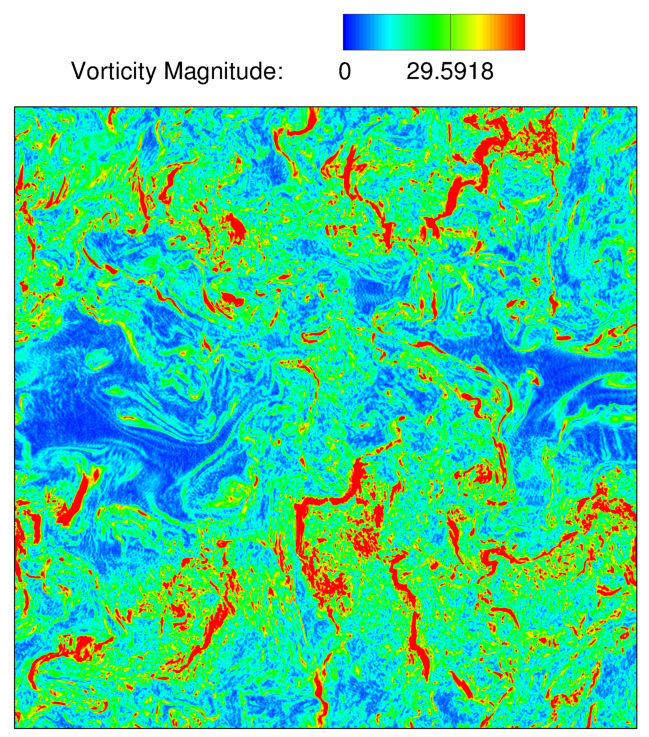}}
\subfigure[]{
\label{con0455}
\includegraphics[scale=0.295]{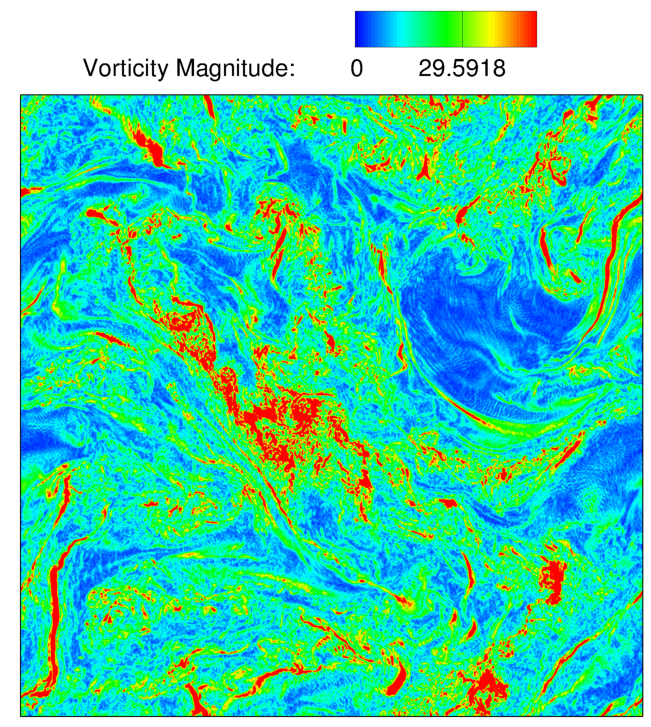}}
\subfigure[]{
\label{fig:stcon0402}
\includegraphics[scale=0.30]{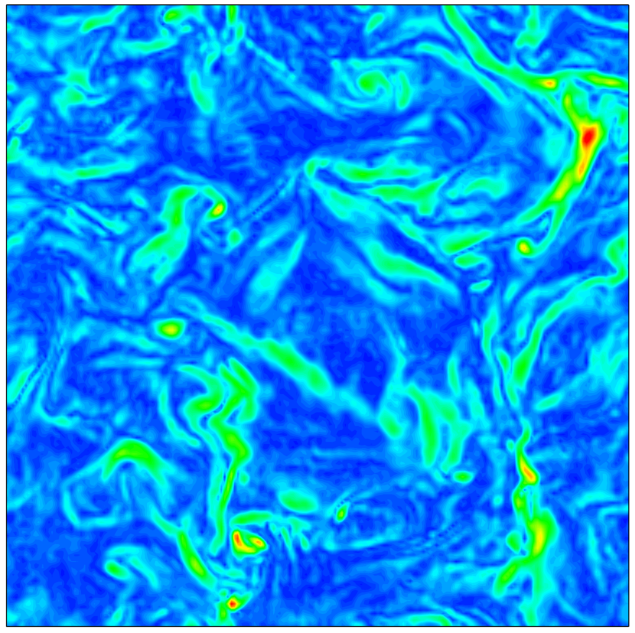}}
\subfigure[]{
\label{stcon0416}
\includegraphics[scale=0.30]{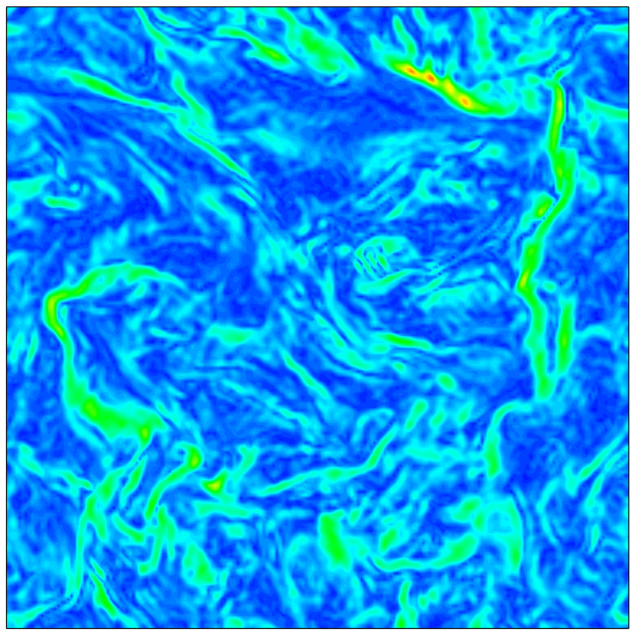}}
\subfigure[]{
\label{stcon0455}
\includegraphics[scale=0.30]{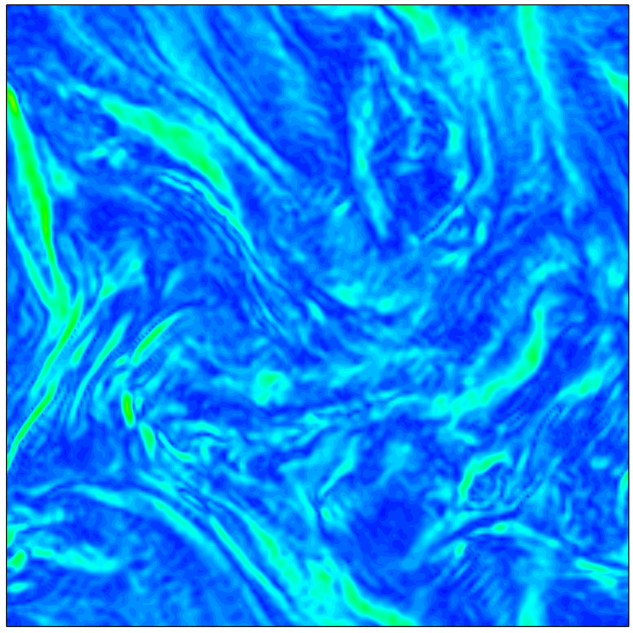}}
\subfigure[]{
\label{dycon0402}
\includegraphics[scale=0.30]{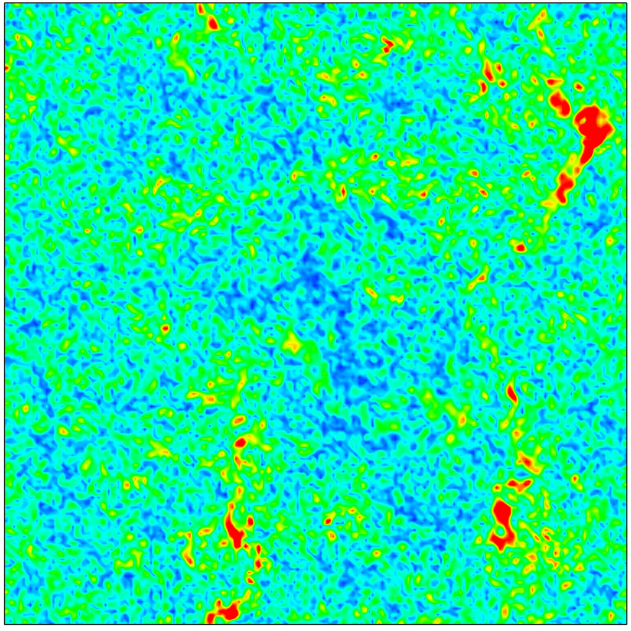}}
\subfigure[]{
\label{dycon0416}
\includegraphics[scale=0.30]{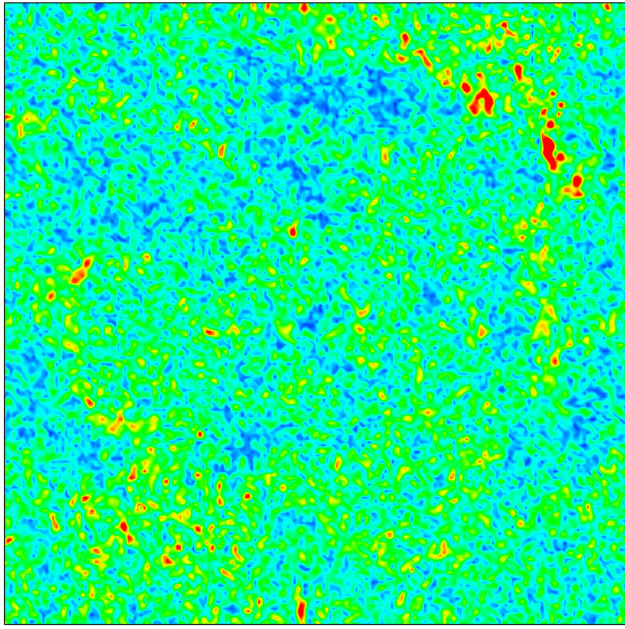}}
\subfigure[]{
\label{dycon0455}
\includegraphics[scale=0.30]{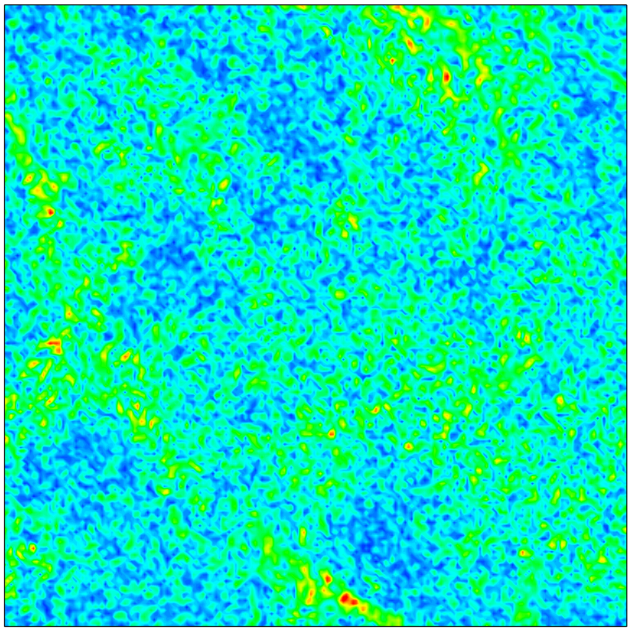}}
\subfigure[]{
\label{nocon0402}
\includegraphics[scale=0.30]{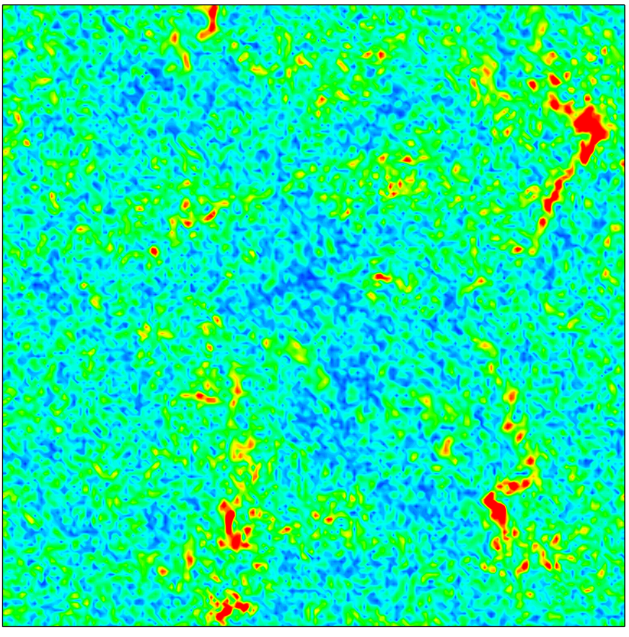}}
\subfigure[]{
\label{nocon0416}
\includegraphics[scale=0.30]{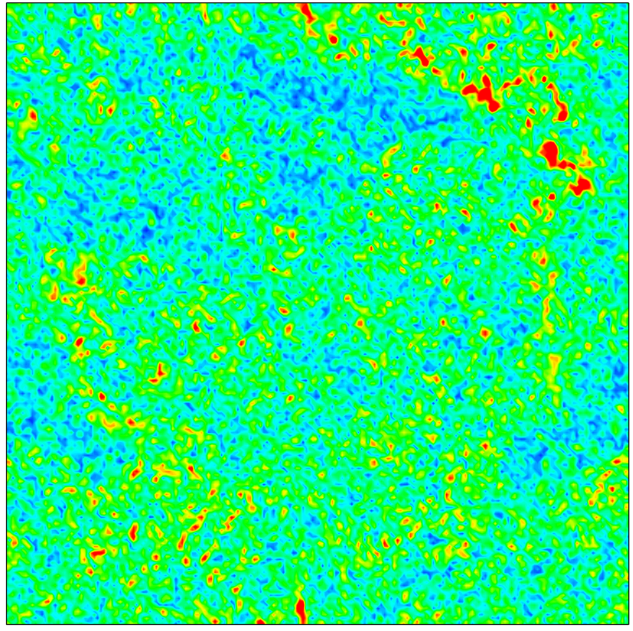}}
\subfigure[]{
\label{nocon0455}
\includegraphics[scale=0.30]{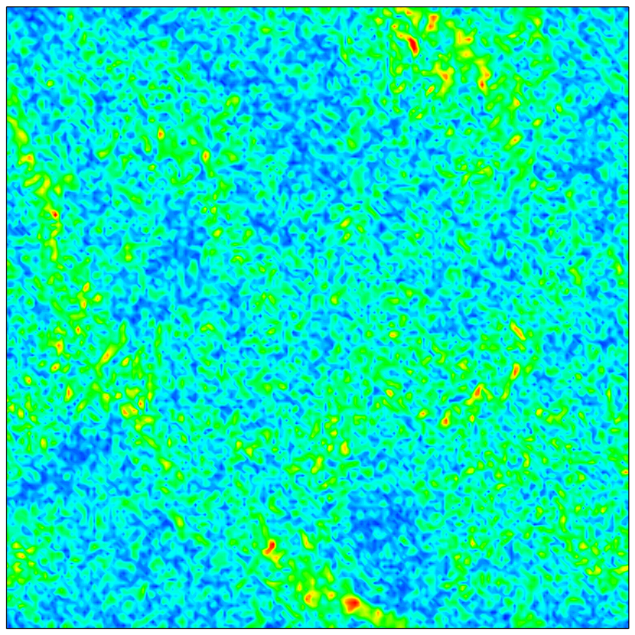}}
\caption{Contours of vorticity magnitude in the $x$-$y$ plane at $z=3.14$ for DNS (a) Case 1 (b) Case 2 (c) Case 3 ; constant coefficient Smagorinsky (CS) model  (d) Case 1 (e) Case 2 (f) Case 3 ; dynamic Smagorinsky (DS) model (g) Case 1 (h) Case 2 (i) Case 3; and dynamic Clark (DC) model (j) Case 1 (k) Case 2 (l) Case 3}
\label{con04}
\end{center}
\end{figure*}

Structurally, it is known that stratification causes the formation of pancake-like horizontal structures due to the transfer of energy from small to large scales. Figure \ref{con04} compares the contour plots of vorticity magnitude for filtered DNS and LES, at the point of maximum dissipation on the $x-y$ plane.  An increase in small structures from Case 1 ($Re_0=3704$) to Case 2 ($Re_0=6667$) as the Reynolds number increases, is observed. Further, large structures are observed due to the domination of stratification over rotation. Also, the inclination of large structures can be seen occurring due to system rotation; similar structural characteristics have been noted previously \cite{agrawal2021large, bin2022evolution}. On comparing the flow structures resulting from LES,  it is apparent that LES employing the constant coefficient Smagorinsky model (Figures \ref{fig:stcon0402},\ref{stcon0416}, \ref{stcon0455}) consists of large scales while finer structures are observed for dynamic Smagorinsky (Figures \ref{dycon0402}, \ref{dycon0416}, \ref{dycon0455}) and dynamic Clark (Figures \ref{nocon0402}, \ref{nocon0416}, \ref{nocon0455}) models. This observation aligns with the respective energy dissipation values observed with these subgrid-scale models (Figure \ref{epsk04}). The lower dissipation values indicate less breakdown of the large scales which is seen from the horizontal length scales in Figure \ref{lh04}. Further, as the stratification strength increases, the large structure emerges due to the dampening of the small scales, thus decreasing the vorticity in the small scales. 

\section{Conclusions}

LES calculations (using a pseudo-spectral code) of decaying stratified-rotating turbulence using three subgrid-scale models, namely: constant coefficient Smagorinsky, the dynamic Smagorinsky, and dynamic mixed Clark models are presented for increasing value of the ratio of the stratification to rotation, or $\mathcal{N}/f$ values of 5, 40 and 138 at Reynolds numbers of 3704, 6667 and 6667, respectively.  Various flow quantities including turbulent dissipation, turbulent kinetic energy, turbulent potential energy, potential and kinetic energy spectra, and horizontal and vertical energy fluxes in the respective directions, are analyzed to understand the capability of the various LES models in predicting the development of rotating stratified turbulence.  

The increase in the oscillation frequency in the evolution of $tke$ and $tpe$ as the stratification strength increases (in comparison to rotation) is exhibited by all LES models. The values of $tke$ and $tpe$ are under-predicted by the constant coefficient Smagorinky model. The two dynamic models over-dissipate kinetic energy compared to the constant coefficient Smagorinsky model. The higher dissipation values result in an under-prediction of the horizontal length scale or $l_h$ as seen in Figure \ref{lh04}, while the lower dissipation for the constant coefficient Smagorinsky model leads to higher values of $l_h$. The energy spectra show a good agreement between DNS and the two dynamic subgrid-scale models for wavenumber $\kappa<20$, with the predictions worsening for large wavenumbers, especially near the LES-cutoff scales. Near the cutoff wavenumber, $\kappa = 32$, the LES calculations exhibit a higher energy content than the DNS suggesting an excessively high energy transfer from large to small scales. It has been shown that the primary reason for this is the energy transfer across horizontal wavenumber. The DNS results indicate a negative or inverse cascade from small to large scales as $\mathcal{N}/f$ increases, which is in contrast to the LES predictions. The kinetic energy spectra start to deviate from the DNS even at the larger scales after $t=3.5$. The evolution of the kinetic energy spectra suggests that the large-scale kinetic energy is transferred towards small scales at later times. Finally, the energy contained in the form of the potential energy is reduced for LES (in comparison to DNS) at later times. 

In summary, both the dynamic subgrid-scale models exhibit higher dissipation values due to the positive transfer of energy from large to small scales as the flow evolves. Yet, they still perform better than the classic constant coefficient subgrid-scale model tested in the current study. Improved subgrid-scale modeling approaches, such as the work of Lund and Novikov \cite{lund1993parameterization}, Agrawal et al. \cite{agrawal2022non} which include the effects of the rotation rate tensor, need to be further validated and analyzed for their use in LES of flows involving phenomena such as rotation and stratification, which are of major interest in the geophysics community.  

\section{Acknowledgements:}

The authors gratefully acknowledge the financial support provided by the Science and Engineering Board (SERB) of the Department of Science and Technology (DST) of the Government of India under the framework of the Startup Research Grant (SRG), No. SRG/2019/000009. Rahul Agrawal acknowledges support from NASA’s Transformational Tools and Technologies project under Grant No. \#80NSSC20M0201 for his graduate studies.

\section{Compliance with Ethical Standards:}
Conflict of Interest: The authors declare that they have no conflict of interest.

\section{Data Availability:}

The data that support the findings of this study are available from the corresponding author upon reasonable request. 

%

\end{document}